# Superconductivity in S-substituted FeTe

Yoshikazu Mizuguchi[1,2,3], Fumiaki Tomioka[1,2], Shunsuke Tsuda[1,2,4], Takahide Yamaguchi[1,2] and Yoshihiko Takano[1,2,3]

1.*National Institute for Materials Science, 1-2-1 Sengen, Tsukuba 305-0047, Japan*

2.*JST, TRIP, 1-2-1 Sengen, Tsukuba 305-0047, Japan*

3.*University of Tsukuba, 1-1-1 Tennodai, Tsukuba 305-0001, Japan*

4.*WPI-MANA-NIMS, 1-1 Namiki, Tsukuba 305-0044, Japan*

Abstract

We synthesized a new superconducting system of $FeTe_{1-x}S_x$ with a PbO-type structure. It has the simplest crystal structure among iron-based superconductors. The superconducting transition temperature is about 10 K when $x$ is 0.2. The upper critical field $B_{c2}(0)$ was estimated to be ~70 T. The coherent length was calculated to be ~2.2 nm. Since $FeTe_{1-x}S_x$ is composed of nontoxic elements, this material is a candidate for applications, which will further advance research on iron-based superconductors.

The discovery of superconductivity in ZrCuSiAs-structured $LaFeAsO_{1-x}F_x$ with a transition temperature $T_c$ = 26 K triggered active studies on iron-based superconductors.[1] The $T_c$ was raised by applying pressure (the onset temperature $T_c^{onset}$ = 43 K) or by the substitution of a smaller rare earth ion for the La site ($T_c$ = 55 K for $SmFeAsO_{1-x}F_x$).[2,3] Various iron-based superconductors, which have a structure analogous to $LaFeAsO_{1-x}F_x$, were discovered. Related compounds include $ThCr_2Si_2$-structured $Ba_{1-x}K_xFe_2As_2$ with $T_c$ = 38 K and $Li_{1-x}FeAs$ with $T_c$ = 18 K.[4,5] Recently, superconductivity at 8 K in PbO-structured FeSe was reported.[6] This compound undergoes a structural phase around 70 K.[7] We reported a huge enhancement of $T_c$ under high pressure (onset temperature $T_c^{onset}$ = 27 K at 1.48 GPa ).[8] The $T_c$ was also raised by the substitution of both S and Te for Se.[9-11] Density functional calculation study indicats that the stability of spin density wave is higher for FeTe than for FeSe, and the doped FeTe achieves a higher $T_c$ than that of FeSe.[12] Although tetragonal FeTe has a structure very analogous to tetragonal FeSe, it does not show a superconducting transition but shows a structural phase transition from tetragonal to orthorhombic around 80 K.[10,13] A FeTe layer, which is composed of nontoxic elements, is very advantageous to applications. We investigated pressure effects of resistivity on $FeTe_{0.92}$ up to 1.6 GPa.[14] With applying pressure, the resistivity decreased and the structural phase transition temperature shifted to a lower temperature, however, superconductivity was not observed down to 2 K. The suppression of the structural phase transition is similar to the reports in other iron-based superconductors that show pressure-induced superconductivity.[15] Thus we believe that the FeTe layer has potential to produce superconductivity if the small element is substituted to induce chemical pressures. Here we report the discovery of superconductivity in S-substituted FeTe.

We prepared $FeTe_{0.92}$ and $FeTe_{1-x}S_x$ ($x$ = 0.1, 0.2) samples using a solid state reaction method. The $FeTe_{1-x}S_x$ samples were synthesized in two ways as follows: (i) The powders of Fe (99.9 %), Te (99.9 %) and S (99 %) were sealed into an evacuated quartz tube with a nominal composition of $FeTe_{1-x}S_x$, and were heated at 800 $^0$C for 12 h. We obtained melted samples with a black surface, which contained single-crystal large grains inside; (ii) We synthesized TeS as a starting material to avoid evaporation of S at low temperatures. Te and S powders were sealed into an evacuated quartz tube with a nominal composition of Te:S = 1:1, and were heated at 400 $^0$C for 12 hours. The appropriate mixture of Fe, TeS and Te powders was sealed into an evacuated quartz tube, and was heated at 600 $^0$C for 12 hours. The obtained samples were reground and pressed into pellets. The pellets were sealed into an evacuated quartz tube again and heated at 600 $^0$C for 12 hours. Finally, we obtained homogeneous black pellets. In this letter, we denote the sample names as the synthesis method, (i) or (ii), and the composition. The obtained samples were characterized by X-ray diffraction using $CuK_{\alpha}$ radiation. The X-ray diffraction pattern for $FeTe_{0.8}S_{0.2}$(ii) was refined by Rietveld refinement using Rietan2000. The ratio of Te:S was also estimated using an energy dispersive X-ray spectrometer (EDX) for $FeTe_{0.8}S_{0.2}$(i). Temperature dependence of resistivity was measured from 300 to 2 K using a four-terminal method. Temperature dependence of magnetization was measured using a SQUID magnetometer down to 2 K.

Figure 1 shows X-ray diffraction patterns for $FeTe_{0.92}$, $FeTe_{0.8}S_{0.2}$(i) and $FeTe_{0.8}S_{0.2}$(ii). The peaks were well indexed using space group *P*4/*nmm*. $FeTe_{0.92}$ sample is a single phase. Weak impurity peaks were observed especially at $2\theta \sim 31.5^{\circ}$ for $FeTe_{0.8}S_{0.2}$(i). For $FeTe_{0.8}S_{0.2}$(ii), the impurity peaks were smaller than those for sample(i). The lattice constants are summarized in Fig. 2. Both *a* and *c* decreased with increasing nominal S content. The *a* and *c* of $FeTe_{0.8}S_{0.2}$(ii) are lager than those of $FeTe_{0.8}S_{0.2}$(i), which implies that the actual S concentration for sample(ii) is smaller than that for sample(i). Figure 3 shows the results of the Rietveld refinement for $FeTe_{0.8}S_{0.2}$(ii) (space group *P*4/*nmm*, $a$ = 3.8123(4) Å, $c$ = 6.2444(8) Å, $V$ = 90.75(2) Å$^3$, $z$ = 0.2790(3), $R_{wp}$ = 14.78 %). The S concentration was estimated to be 5.8 % of Te. The obtained S concentration for $FeTe_{0.8}S_{0.2}$(ii) is about 1/4 of the starting nominal composition. We also estimated the ratio of Te:S using EDX for $FeTe_{0.8}S_{0.2}$(i). The ratio of Te:S was estimated to be 90:10 for the outer surface of $FeTe_{0.8}S_{0.2}$(i) and 96:4 for the inner of $FeTe_{0.8}S_{0.2}$(i). The actual S concentrations estimated from Rietveld refinement and EDX are smaller than the starting nominal composition. Due to the difference of the ionic radius between S and Te, the solid solubility limit of S for the Te site may be low for the synthesis at ambient pressure.

Figure 4 (a) shows the temperature dependence of resistivity for $FeTe_{0.92}$, $FeTe_{0.9}S_{0.1}$(i) and $FeTe_{0.8}S_{0.2}$(i). Fig. 4 (b) shows an enlargement of the low temperature range. For $FeTe_{0.92}$, a strong anomaly, which corresponds to a structural phase transition, was observed around 80 K. The structural phase transition temperature shifted below 50 K for $FeTe_{0.9}S_{0.1}$(i) and disappeared for $FeTe_{0.8}S_{0.2}$(i). With increasing S concentration, the normal-state resistivity decreased. At low temperatures, superconducting transitions were clearly observed for the S-substituted samples. The $T_c^{onset}$s were estimated to be 10.0 and 10.5 K for $FeTe_{0.9}S_{0.1}$(i) and $FeTe_{0.8}S_{0.2}$(i) from a point at which the resistivity deviates from the linear temperature dependence. A zero resistivity was observed at 7.8 K for $FeTe_{0.8}S_{0.2}$(i). Fig. 5 (a) shows the temperature dependence of resistivity for $FeTe_{0.8}S_{0.2}$(i) under magnetic fields up to 7 T. The estimated $T_c^{onset}$, the midpoint temperature $T_c^{mid}$ and the zero resistivity temperature $T_c^{zero}$ were plotted as a function of magnetic field in Fig. 5 (b). We estimated the upper critical field $B_{c2}(0)$ and the irreversible field $B_{irr}(0)$ to be 102 and 56 T from the linear extrapolation in Fig. 5 (b). Assuming that this superconductivity is in the dirty limit,[16] $B_{c2}(0)$ is estimated to be ~ 70 T. From the $B_{c2}$, the coherent length was estimated to be ~2.2 nm using the equation of $\xi^2=\Phi_0/2\pi B_{c2}$. Fig. 5 (c) shows the temperature dependence of magnetization for $FeTe_{0.9}S_{0.1}$(i) and $FeTe_{0.8}S_{0.2}$(i) with zero field cooling (ZFC) and field cooling (FC) modes. The $T_c^{mag}$ ($T_c$ estimated from the magnetization) was found to be 8.6 K for both samples. With increasing S concentration, the shielding volume fraction was considerably enhanced. Figure 6 (a) shows the temperature dependence of resistivity for $FeTe_{0.92}$ and $FeTe_{0.8}S_{0.2}$(ii), and Fig. 6 (b) shows an enlargement of the low temperature range. The S substitution induced superconductivity and suppressed the structural phase transition around 80 K observed in the mother phase. The $T_c^{onset}$ and $T_c^{zero}$ were estimated to be 8.8 and 2.8 K.

The lattice constants clearly decreased with increasing S concentration. The S substitution corresponds to a positive chemical pressure. The decrease of resistivity and the suppression of the structural phase transition with S substitution are in good agreement with the results in pressurized FeTe.[14] The suppression of the structural phase transition should be the key to produce superconductivity in FeTe. These results imply the possibility of superconductivity in FeTe under high pressure.

We synthesized a new superconducting system of $FeTe_{1-x}S_x$. The superconductivity is observed in the FeTe layer. To date, superconductivity has been reported only in the FeP, FeAs or FeSe layers for iron-based superconductors. FeTe-based superconductor is advantageous for application, because it is composed of nontoxic elements. Furthermore, $FeTe_{1-x}S_x$ has a very high $B_{c2}(0)$ as high as 70 T. Therefore, FeTe-based superconductor is a candidate material for applications. The search for new FeTe-based superconductors that have a structure such as ZrCuSiAs-type, $ThCr_2Si_2$–type or multi stacking one, will be an appealing stage.

This work was partly supported by Grant-in-Aid for Scientific Research (KAKENHI).

Figure captions

Fig.1. X-ray diffraction pattern for $FeTe_{092}$, $FeTe_{0.8}S_{0.2}$(i) and $FeTe_{0.8}S_{0.2}$(ii). The unidentified peaks are indicated with asterisks.

Fig.2. Lattice constants (a)$a$ and (b)$c$ for all samples are plotted as a function of $x$.

Fig.3. Results of Rietveld refinement for $FeTe_{0.8}S_{0.2}$(ii).

Fig.4. (a) Temperature dependence of resistivity for $FeTe_{0.92}$, $FeTe_{0.9}S_{0.1}$(i) and $FeTe_{0.9}S_{0.1}$(i). Arrows in this figure indicate the maximum point of the anomaly. (b) An enlargement of the low temperature range.

Fig.5. (a) Temperature dependence of resistivity for $FeTe_{0.8}S_{0.2}$(i) under magnetic fields up to 7 T with an increment of 1 T. (b) $B$-$T_c$ plot. The $B_{c2}$ and the $B_{irr}$ were estimated from the linear extrapolation. (c) Temperature dependence of magnetization for $FeTe_{0.9}S_{0.1}$(i) and $FeTe_{0.8}S_{0.2}$(i).

Fig.6 (a) Temperature dependence of resistivity for $FeTe_{0.92}$ and $FeTe_{0.8}S_{0.2}$(ii). (b) An enlargement of the low temperature range.

Fig.1

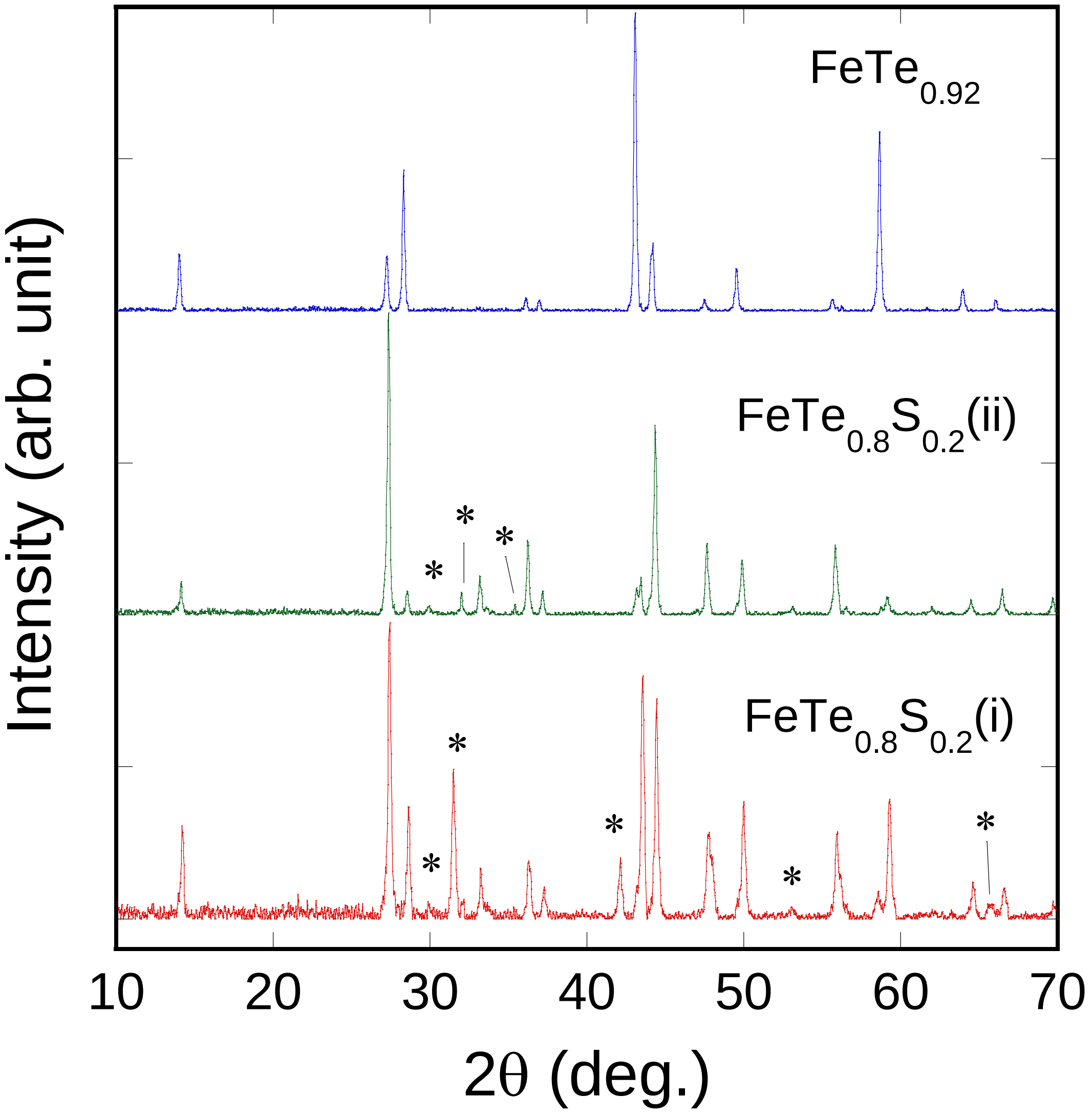

Fig.2

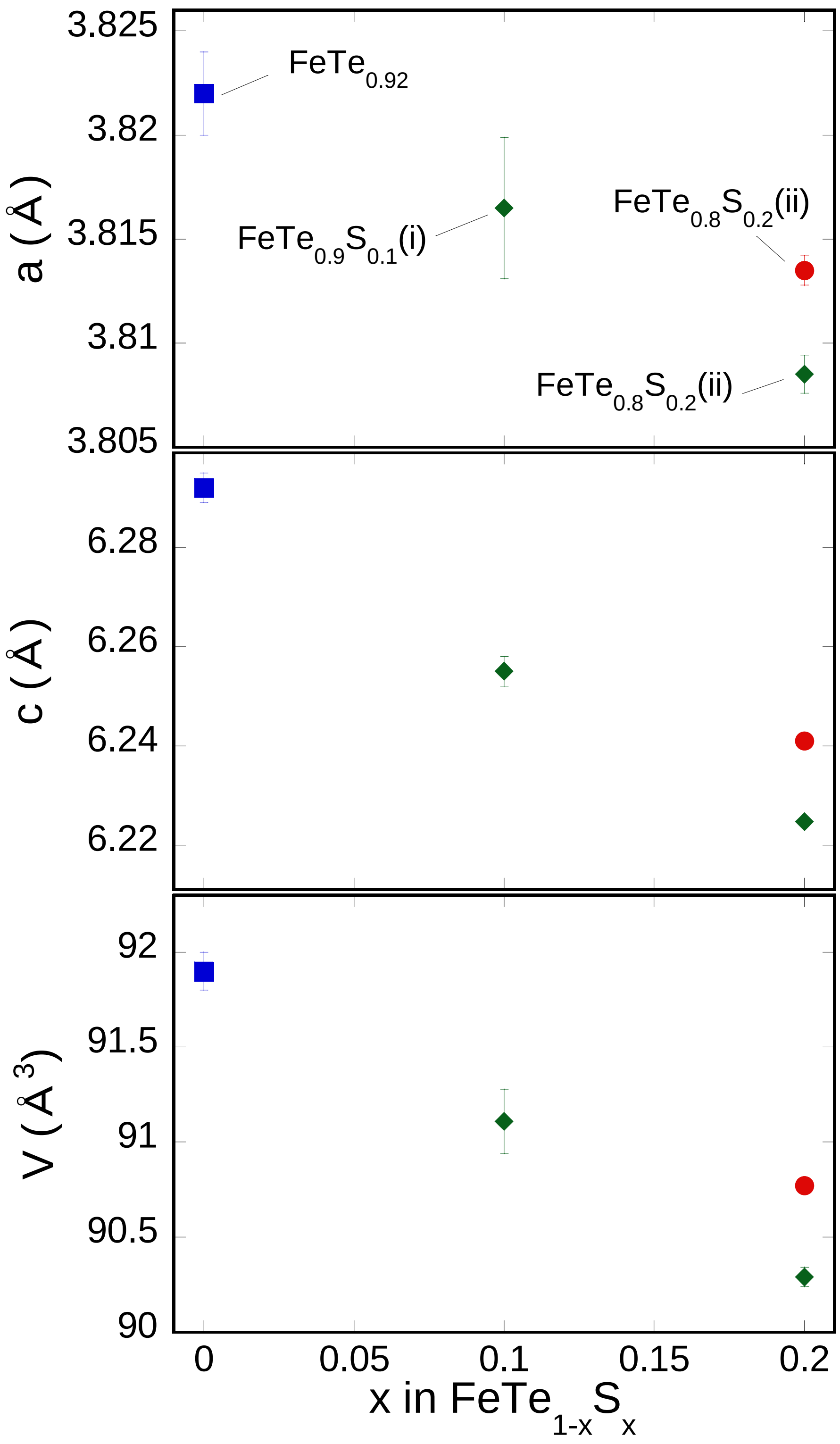

Fig.3

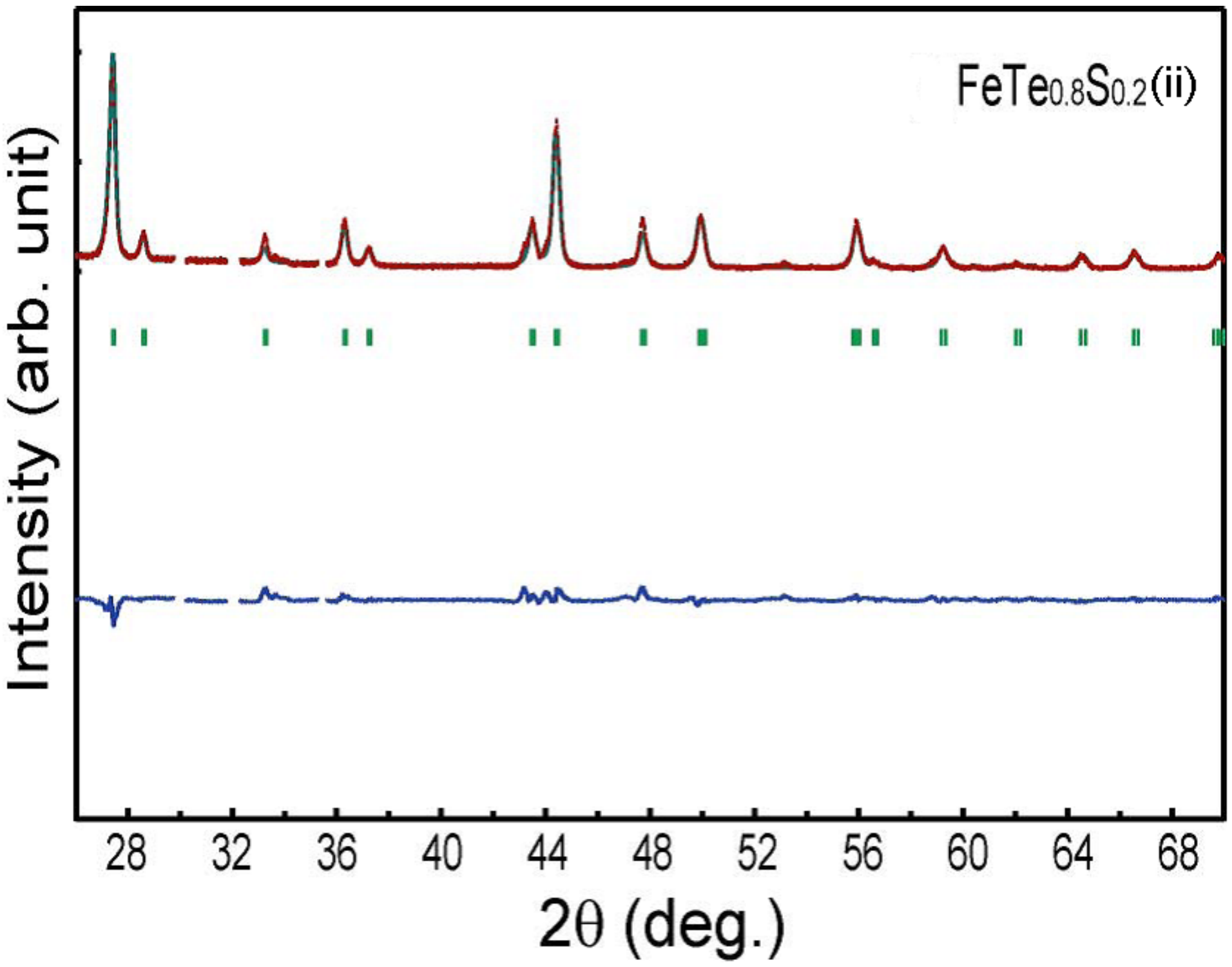

Fig.4 (a)

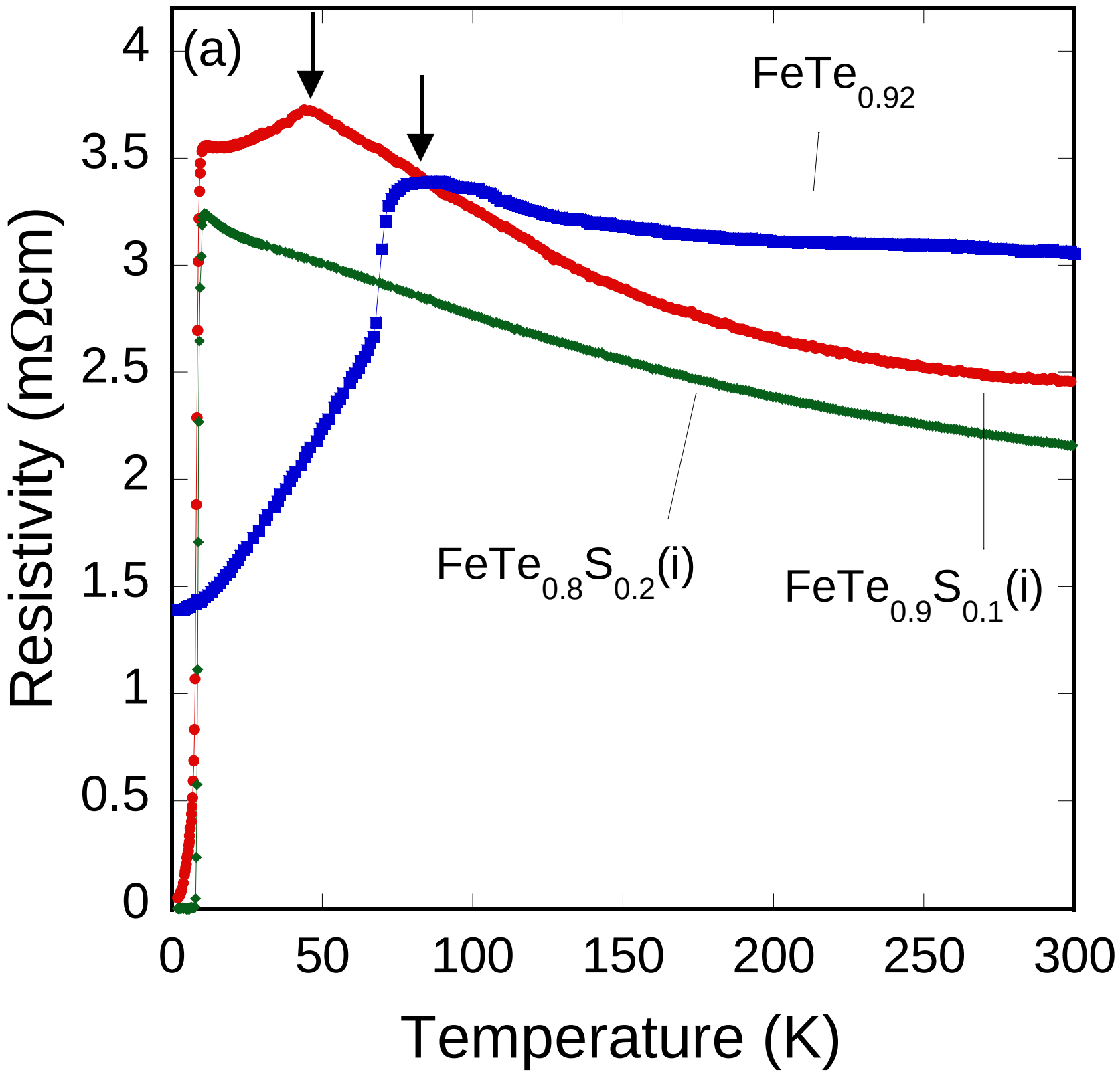

(b)

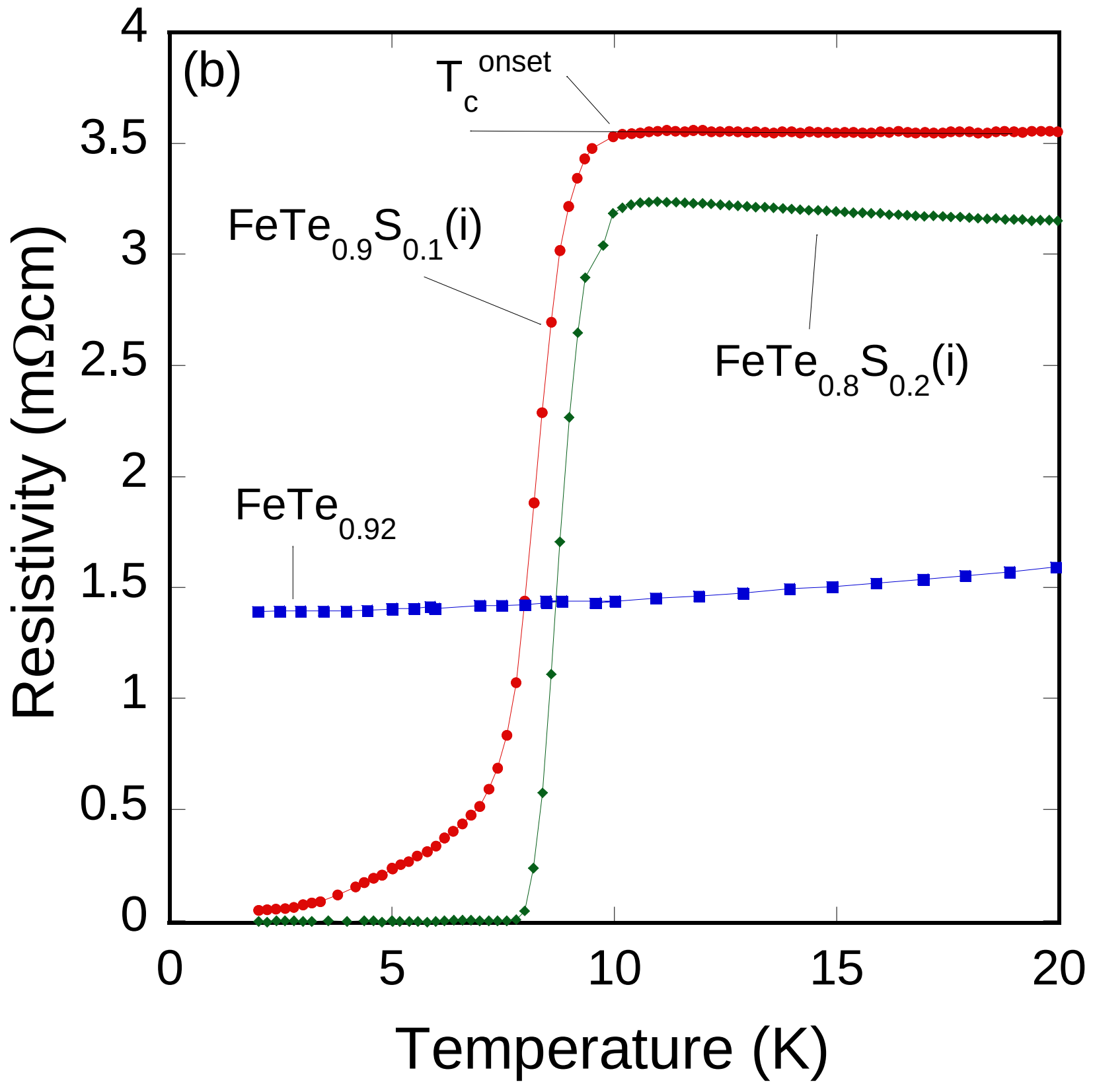

Fig.5 (a)

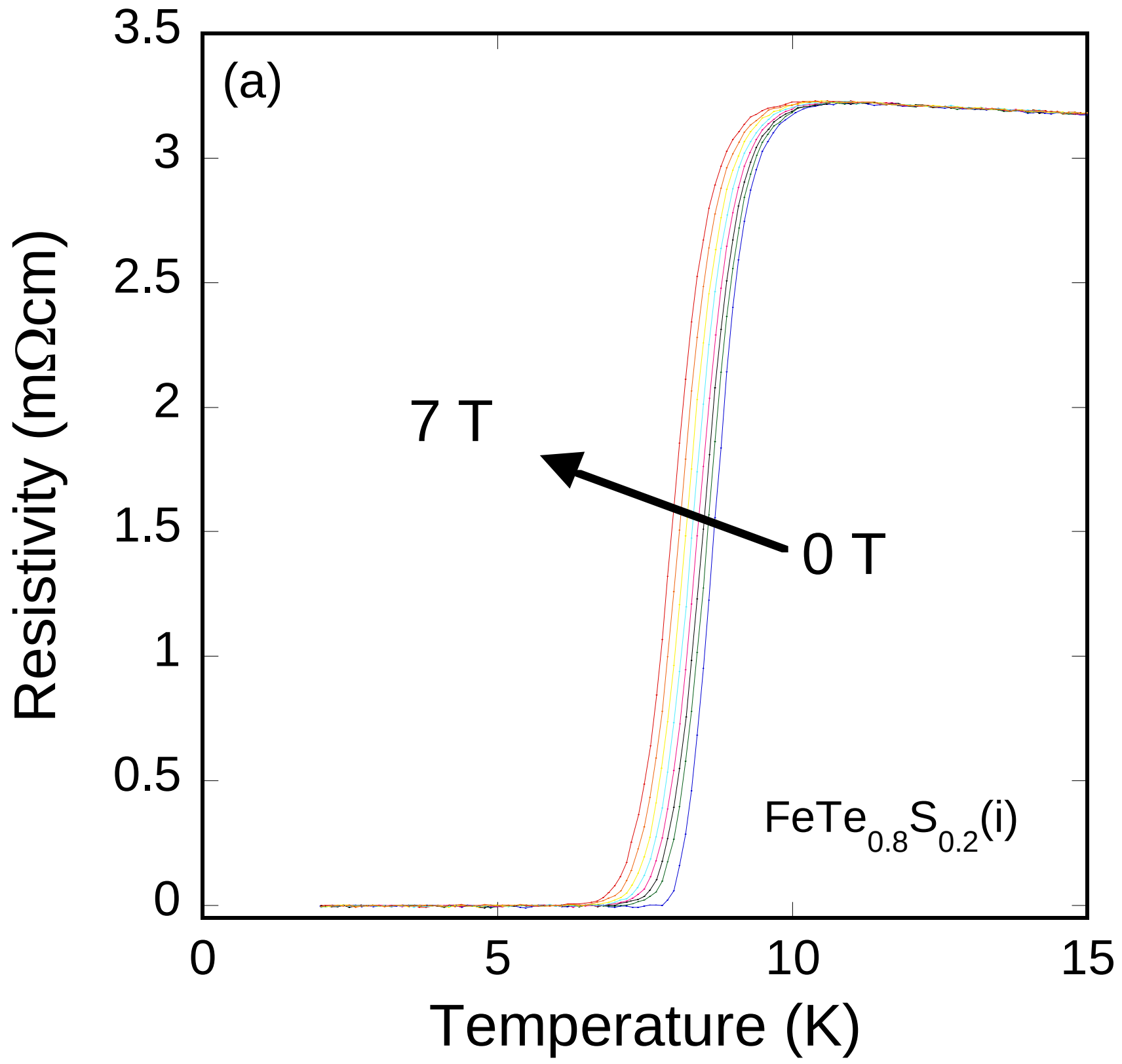

Fig.5 (b)

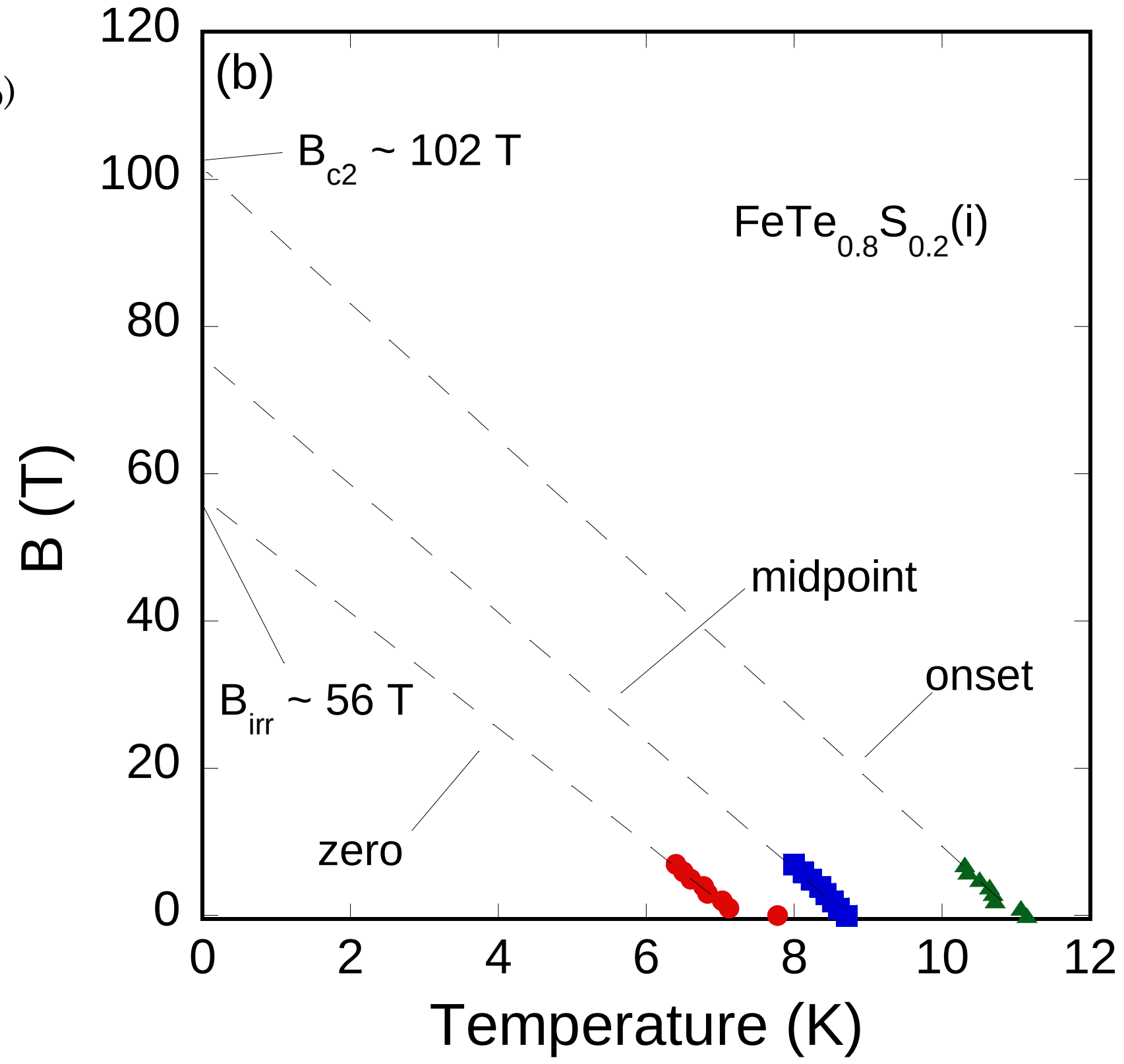

Fig.5 (c)

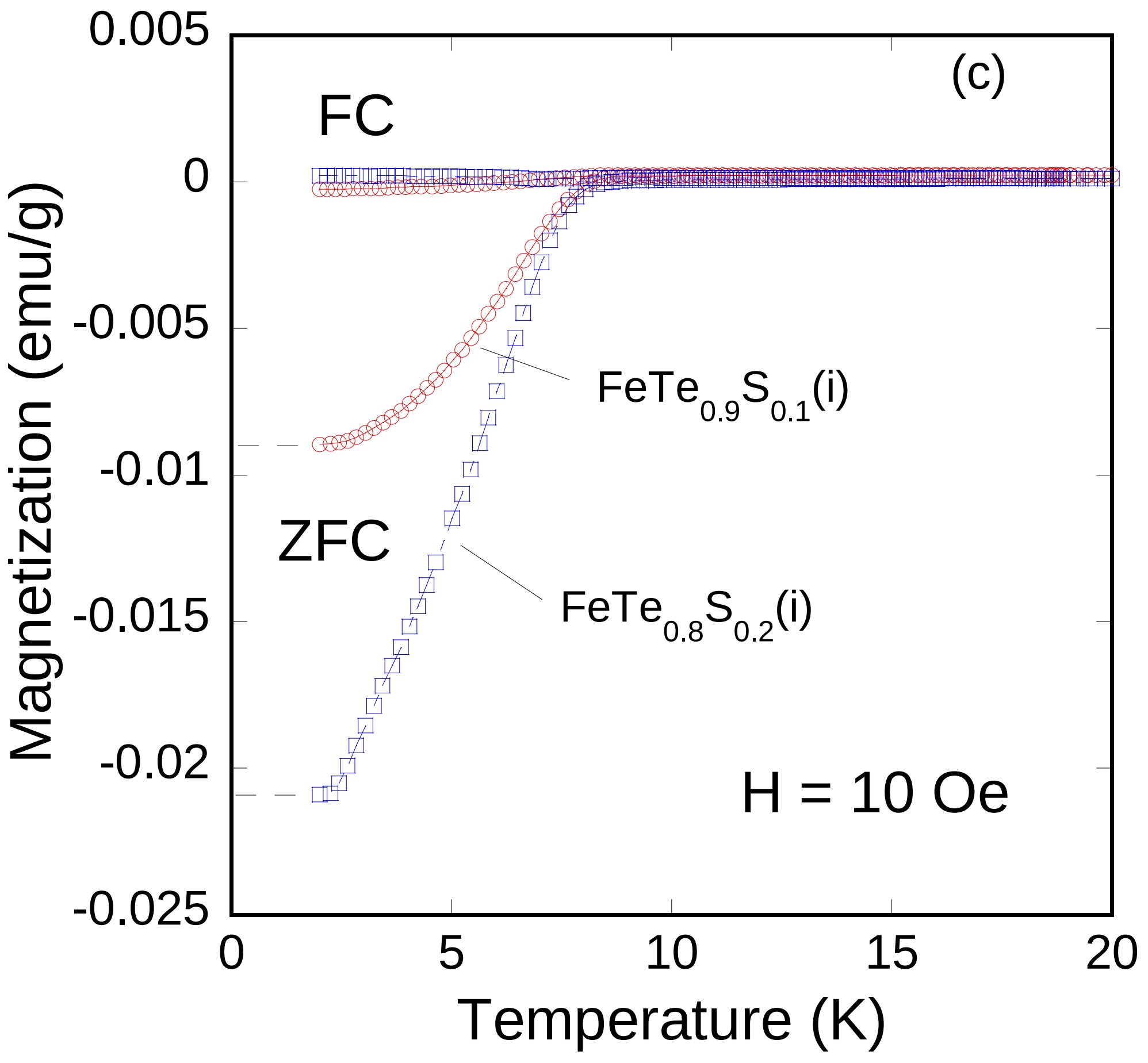

Fig.6 (a)

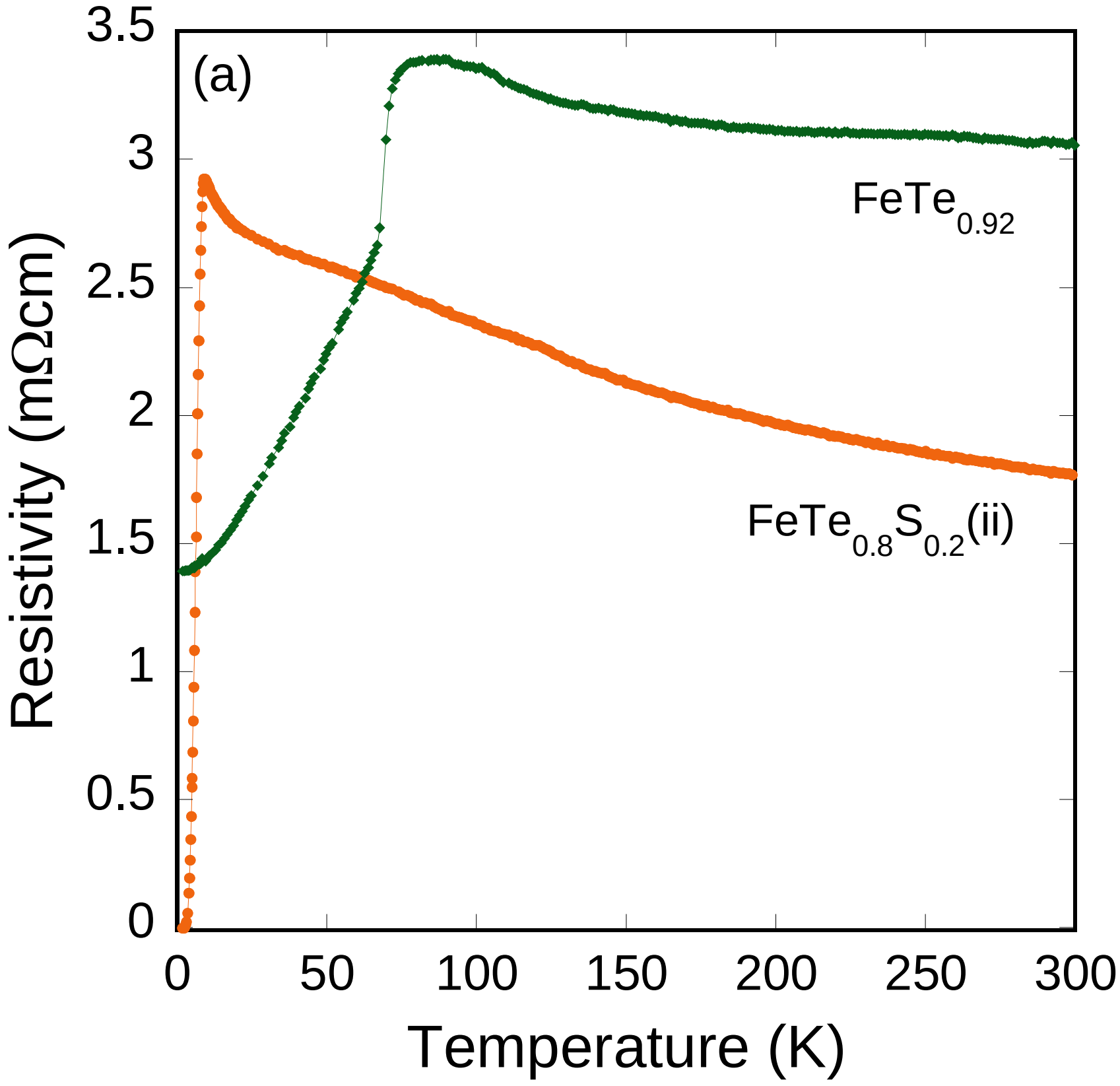

Fig.6 (b)

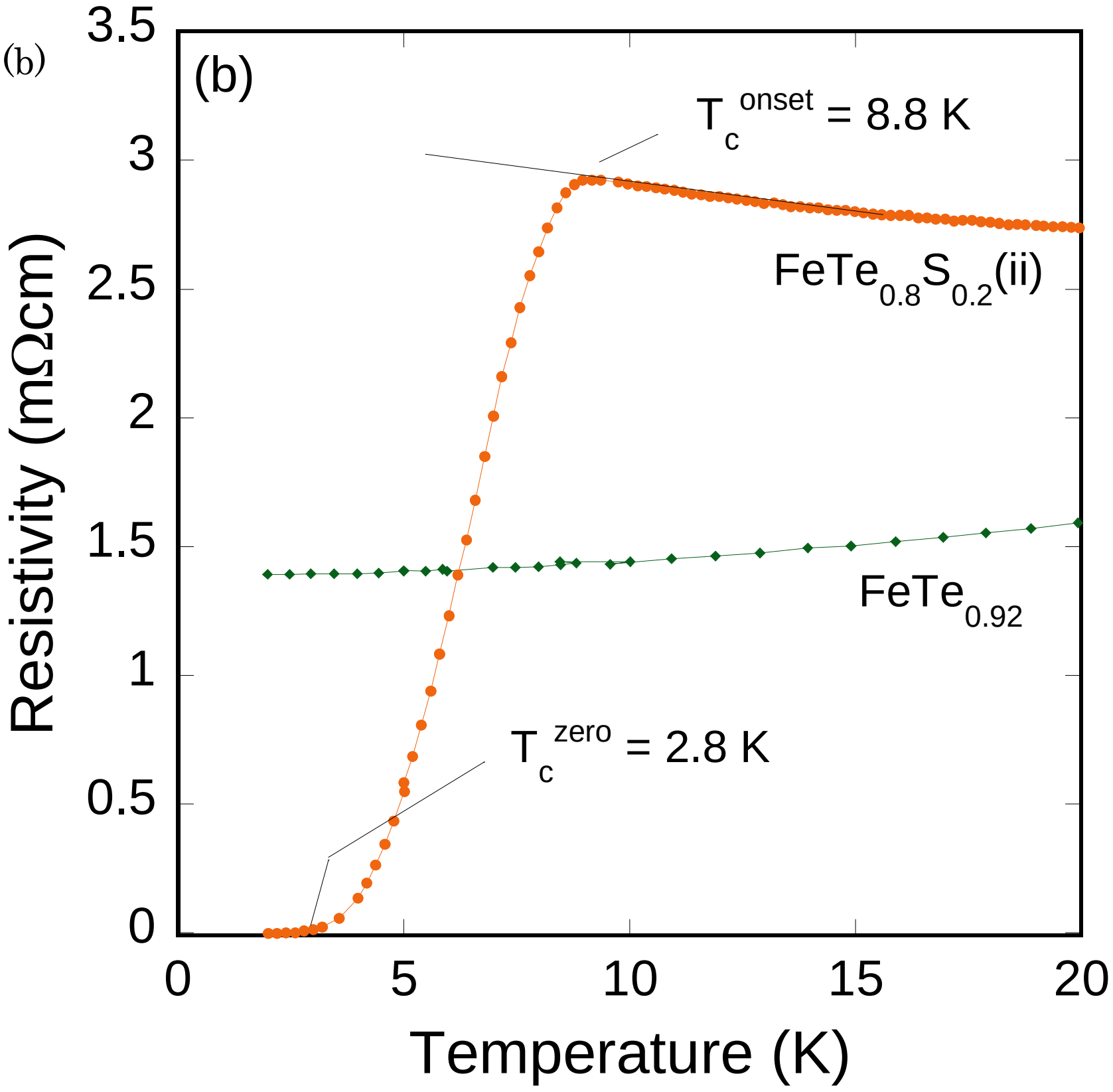